%% file: di-jetPaper.tex
\documentclass[aps, prl, reprint, superscriptaddress, floatfix, showpacs, amsmath, amssymb]{revtex4-1}

\usepackage{graphicx}
\usepackage{dcolumn}
\usepackage{bm}

\newcommand {\MM} [1] {\ensuremath{#1}}

\newcommand {\IT} [1] {\ensuremath{#1}}

\newcommand {\SQRT}[1] {\ensuremath{\sqrt{#1}}}

\newcommand {\SUB} [2] {\MM{#1\ensuremath{_{#2}}}}
\newcommand {\SUP} [2] {\MM{#1\ensuremath{^{#2}}}}

\newcommand {\proton} {\IT{p}}

\newcommand {\capsword} [1] {\textsc{#1}}
\newcommand {\NLO} {\capsword{NLO}}
\newcommand {\QCD} {\capsword{QCD}}

\newcommand {\RHIC} {\capsword{RHIC}}

\newcommand {\STAR} {\capsword{STAR}}

\newcommand {\BEMC} {\capsword{BEMC}}
\newcommand {\EEMC} {\capsword{EEMC}}

\newcommand {\TPC} {\capsword{TPC}}

\newcommand {\pp} {\proton\proton}

\newcommand {\Dgx} {\IT {\Delta g(x)}}
\newcommand {\ALL} {\SUB{\IT A}{\IT{LL}}}

\newcommand {\invpb} {\SUP{\rm pb}{-1}}

\begin{document}

\title{\texorpdfstring{Measurement of the cross section and
longitudinal double-spin asymmetry \\
for di-jet production in polarized
{\boldmath $pp$} collisions at {\boldmath $\sqrt{s}$} = 200~GeV}
{Title}}

\include{STAR-Authors-Feb2017}




\date{\today}

\begin{abstract} 
We report the first measurement of the longitudinal double-spin asymmetry
\ALL\ for mid-rapidity di-jet production in polarized \pp\
collisions at a center-of-mass energy of \SQRT{s}\ = 200~GeV. The di-jet
cross section was measured and is shown to be consistent with
next-to-leading order (\NLO) perturbative \QCD\ predictions. \ALL\
results are presented for two distinct 
topologies, defined by the jet
pseudorapidities, and are compared to predictions from several recent
\NLO\ global analyses. The measured asymmetries, the first such correlation 
measurements, support those analyses
that find positive gluon polarization at the level of roughly 0.2 over the region of Bjorken-$x > 0.05$.
\end{abstract}

\maketitle


Determining the helicity distribution of the gluons within a proton as a function of momentum fraction,
\Dgx, remains an important challenge in high-energy
nuclear physics. We do not yet understand the
decomposition of the proton's spin into contributions from the
spins and orbital angular momenta of its internal quarks and gluons,
although high-precision, polarized deep-inelastic scattering
(DIS) experiments \cite{Aidala:2012mv} 
have shown that less than a third is due
to the summed intrinsic spins of the quarks and anti-quarks for $x \gtrsim 10^{-3}$
\cite{deFlorian:2008mr,Blumlein:2010rn,Leader:2010rb,Ball:2013lla}. 
These fixed-target polarized DIS data only weakly constrain the gluon 
polarization from inclusive measurements through scaling violations due 
to the limited coverage of photon virtuality $Q^2$.

The Relativistic Heavy Ion Collider (\RHIC) has enabled more direct 
studies of gluons by colliding beams of high-energy polarized
protons \cite{Alekseev:2003sk}, which directly involve gluons 
via the quark-gluon $(qg)$ and gluon-gluon $(gg)$ scattering processes 
that dominate at RHIC $pp$ 
energies \cite{deFlorian:1998qp}. While 
leading-order analyses of DIS data with high-$p_T$ hadron pairs have shown 
hints of positive gluon polarization 
\cite{Adolph:2012vj,Airapetian:2010ac},
the tightest constraints on \Dgx\ 
and its integral over moderate gluon momentum fractions, $x > 0.05$, 
are provided by next-to-leading-order (\NLO) perturbative QCD (pQCD) 
global 
analyses that incorporate the inclusive jet 
\cite{Abelev:2006uq,Abelev:2007vt,Adamczyk:2012qj,Adamczyk:2015}
and $\pi^0$ \cite{Adare:2008aa,Adare:2008qb,Adare:2014hsq}
longitudinal double-spin asymmetries measured by STAR
and PHENIX, respectively, at RHIC.  The most recent such analyses 
\cite{deFlorian:2014yva,Nocera:2014gqa} now
find compelling evidence for positive gluon polarization of roughly 0.2 over the
range $x > 0.05$; they also demonstrate the importance of the RHIC data 
in reaching this conclusion.

Inclusive jet and $\pi^0$ measurements, however, necessarily integrate over a large range in $x$ 
of the initial state partons for a given transverse momentum, $p_T$, of the final state. To
gain more direct sensitivity to the $x$ dependence of $\Delta g$, correlation
measurements, such as di-jet production, are required, as these more 
tightly 
constrain the kinematics of the colliding partons. At leading order in 
QCD, the di-jet invariant mass is proportional to the square-root of the product of the initial 
state momentum fractions, $M =\sqrt{s}\sqrt{x_1 x_2}$, while the sum of the 
jet pseudorapidities determines their ratio, $\eta_1 + \eta_2 = 
\ln(x_1/x_2)$.

In this letter, we report the cross section as well as the first measurement of the 
longitudinal double-spin asymmetry, \ALL, for di-jets produced in 
longitudinally
polarized $\overrightarrow{p} + \overrightarrow{p}$ collisions at
\SQRT{s}\ = 200~GeV, based on data recorded in 2009 by the STAR
collaboration. The asymmetry result was obtained from a data set of 
integrated luminosity 21~\invpb; the
cross section is based on a 19~\invpb\ subset of these data. The
polarization of each of the two colliding proton beams, denoted blue (B)
and yellow (Y), was determined for each RHIC fill using proton-carbon-based
Coulomb-Nuclear Interference polarimeters \cite{Jinnouchi:2004up}, which
were calibrated using a polarized hydrogen gas-jet target 
\cite{Okada:2005gu}. The
luminosity-weighted average polarizations of the two beams were $P_B =
56\%$ and $P_Y = 57\%$. The $A_{LL}$ analysis took into account 
the decay of beam polarization over the course of a RHIC fill. The 
product $P_{B} \, P_{Y}$ used in the 
asymmetry measurement had a relative uncertainty of 6.5\% \cite{CNI:2009}.

The \STAR\ detector subsystems used to reconstruct jets
are the Time Projection Chamber (\TPC) and the Barrel and
Endcap Electromagnetic Calorimeters (\BEMC, \EEMC)
\cite{NIM-RHIC}. The \TPC\ provides charged particle tracking in a 0.5~T
solenoidal magnetic field over the range $|\eta| \lesssim 1.3$ in 
pseudorapidity and $2\pi$ in azimuthal angle $\phi$. The \BEMC\ and 
\EEMC\ are
segmented lead-scintillator sampling calorimeters, which provide full
azimuthal coverage for $|\eta| < 1$ and $1.09 < \eta < 2$,
respectively. The calorimeters measure electromagnetic energy deposition
and provide the primary triggering information via fixed $\Delta \eta \times \Delta \phi = 1 \times 1$ calorimeter regions called jet patches. A jet patch trigger was 
satisfied if the transverse energy in a single jet patch exceeded either 
5.4~GeV (JP1 trigger) or 7.3~GeV (JP2 trigger), or if two  
jet patches adjacent in azimuth each exceeded 3.5~GeV (AJP trigger). 
Details of the track momentum, and 
calorimeter energy resolutions can be found in \cite{Adamczyk:2012qj}. In
addition, the Beam-Beam Counters (BBCs) \cite{BBC:2005} were used in the 
determination of the integrated 
luminosity and, along with the zero-degree calorimeters 
(ZDCs) \cite{NIM-RHIC}, in the determination of helicity-dependent 
relative luminosities. 

The jet reconstruction procedures for these analyses follow those used in 
the inclusive jet analysis from 2009 \cite{Adamczyk:2015}. Jets were found 
using the anti-$k_{T}$ algorithm \cite{ANTIKT} as implemented in the 
FastJet \cite{FASTJET} package, using charged-particle track 
momenta from the \TPC\ and electromagnetic energy from the calorimeters 
as inputs. The resolution parameter $R = 0.6$ sets the effective size of the 
jet in $\eta$-$\phi$ space. To be included in the jet analysis, 
individual tracks were required to have a $p_{T} \ge 0.2$~GeV/$c$ and 
individual calorimeter towers
needed $E_T \ge 0.2$~GeV. 
To avoid double-counting jet energy contributions 
from the \TPC\ and calorimeters, towers with tracks pointing to them had the 
corresponding track energy $p_{T}c$ subtracted from the $E_{T}$ of the tower, 
then negative energies were set to zero. This method resutlts in 
a jet energy resolution of 18\% \cite{Adamczyk:2015}.

Di-jets were selected by 
choosing the two jets with the highest $p_{T}$ from a 
single event that fell in the pseudorapidity range 
$-0.8 \leq \eta \leq 0.8$. These jets 
were required to be more than 120$^\circ$ apart in azimuth. Further 
conditions were placed on the di-jets in order to ensure they reflected the 
partonic hard scattering and to reduce the contributions from background. 
This required that at least one jet contained energy from charged tracks, and 
di-jets containing tracks with $p_{T}$ above 30~GeV/c, where TPC momentum 
resolution is poor, were removed from the analysis. The later cut 
was implemented to eliminate di-jets with highly imbalanced jet transverse 
momenta. These events arose when one track in the event was mis-reconstructed 
to have an artificially high $p_{T}$. To 
facilitate comparison with theoretical predictions, an asymmetric condition 
was placed on the transverse momenta of the jets \cite{deFlorian:1998qp}, 
such that one jet in the 
pair had $p_{T} \geq 8.0$~GeV/$c$ and the other had $p_{T} \geq 
6.0$~GeV/$c$. 
Finally, it was required that at least one jet in the pair points to a 
jet patch that fired the JP2 or AJP (asymmetry and cross section) or JP1 
(asymmetry only) trigger.  

To correct for detector effects on the measured jet quantities and to
estimate systematic uncertainties, simulated events were created using 
PYTHIA 6.425 \cite{Sjostrand:2006za} with the Perugia 0 tune 
\cite{Perugia_ref} and run through a STAR detector response package 
implemented in GEANT 3 \citep{GEANT}. The simulated events were embedded 
into `zero-bias' data events, which were 
triggered on random bunch crossings
over the span of the run, allowing the 
simulation sample to account properly for the beam background, pile-up,
and detector conditions
seen in the data set.

\begin{figure}[t!]
\centering \includegraphics[keepaspectratio=true,
width=0.99\linewidth]{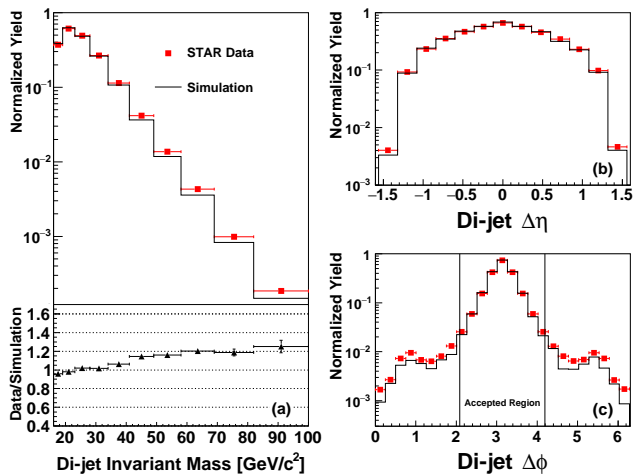}
\caption[Data Simulation Comparison]{(Color online) Comparison of di-jet yields as a function of di-jet invariant mass (a), pseudorapidity gap (b), and azimuthal opening angle between the jets (c) in data and Monte Carlo. The distributions in (a) and (b) are taken from events within the accepted $\Delta \phi$ region shown in (c).}
\label{fig:DATASIMU}
\end{figure} 

Detector-level di-jets were reconstructed from the simulated TPC and 
calorimeter responses using the same jet-finding algorithm as for the data. 
Figure 
\ref{fig:DATASIMU} compares the distributions of the di-jet invariant mass,  
as well as the pseudorapidity gap and azimuthal opening angle between the two jets, for di-jets reconstructed from data and 
simulation, and confirms that the STAR detector response is well understood.
Di-jets were also reconstructed in simulation at the 
particle and parton levels using the anti-$k_{T}$ algorithm. 
Particle-level di-jets were formed from stable, 
final-state particles produced in the simulated event, while parton-level 
di-jets were reconstructed from the hard-scattered partons emitted in the 
collision, including initial and final-state radiation, but not  
beam remnants or underlying event effects as discussed below.

The differential di-jet cross section was calculated at the particle level as a 
function of invariant mass and pseudorapidity according to 
\begin{equation} \label{eq:CROSSSEC}
\frac{\mathrm{d}^{3}\sigma}{\mathrm{d} M \mathrm{d}\eta_{1}\mathrm{d}\eta_{2}} =
\frac{1}{\Delta M \Delta \eta_{1} \Delta \eta_{2}}
\frac{\mathrm{J}}{\mathcal{L}} ,
\end{equation}

\noindent where $\Delta M$ and $\Delta \eta$ are the invariant 
mass and jet pseudorapidity intervals, $\mathcal{L}$ is the integrated 
luminosity of the sample, and J is the fully corrected di-jet yield. 
The corrected yield was obtained by unfolding the raw di-jet yield to 
the particle level using the Singular Value 
Decomposition (SVD) method as implemented within the RooUnfold package 
\cite{Adye:2011gm}, which corrects for bin migration effects due to finite 
detector resolution and acceptance. The input to SVD is a simulated `response matrix', 
which relates the mass of di-jets found at the detector level to the mass 
of the corresponding di-jets at particle level on an event-by-event basis. 
Di-jet matching between detector and 
particle level was done by finding the closest particle-level jet in 
$\eta$-$\phi$ space to each detector-level jet in the event, and requiring 
these jets to be within $\sqrt{\Delta \eta^2 + \Delta \phi^2} \leq 0.5$. 
There is a modest but systematic tendency for the detector-level 
di-jet mass to fall below the particle-level mass due to finite track 
reconstruction efficiency. The mass migration and detector-level purities 
encoded in the 
response matrix are given in the supplemental materials \cite{Supplement}.
Once 
the raw yield had been unfolded back to the particle level, a correction 
for the detector, reconstruction, and trigger efficiencies was applied.

\begin{figure}[t!]
\centering \includegraphics[keepaspectratio=true,
width=0.99\linewidth]{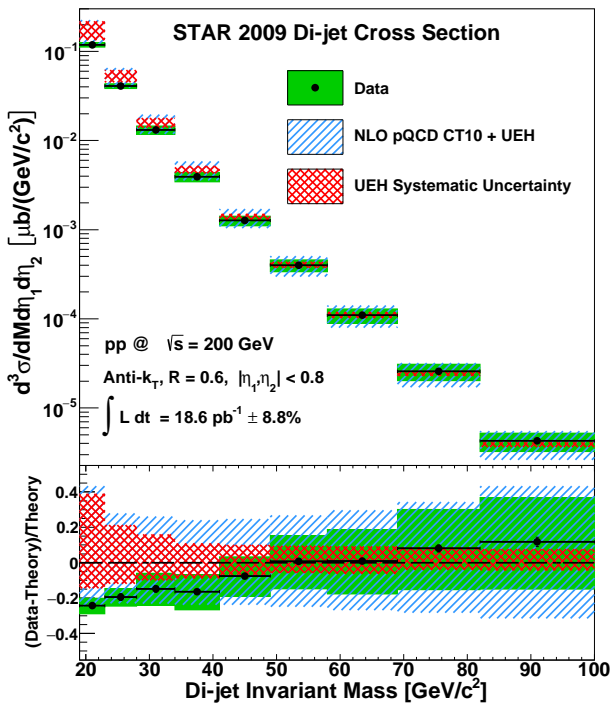}
\caption[Cross Section]{(Color online) The particle-level di-jet differential cross section measured by the STAR experiment (points plotted at bin center). The lower panel provides a relative comparison to theory, as
described in the text.}
\label{fig:XSEC}
\end{figure} 

Figure \ref{fig:XSEC} shows the measured di-jet cross section, indicating 
the associated systematic uncertainty (solid green band) and a 
theoretical prediction 
(single-hatched blue bar) obtained from the NLO di-jet production code of 
de Florian {\it et al.} \cite{deFlorian:1998qp} using the CT10 parton 
distribution function (PDF) set 
\cite{Lai:2010vv} (see supplementary material for values \cite{Supplement}). The 
systematic uncertainty budget of the data contains 
contributions from uncertainties on track reconstruction efficiency and 
calorimeter tower energy scale (each ranging
from 3\% to 15\%) as well as uncertainties on track $p_{T}$ resolution and 
the unfolding procedure. The detector uncertainties were 
propagated to the cross section by altering the simulated detector 
response when creating the response matrix, and then using this modified 
matrix to extract a new cross section. The systematic uncertainty is the 
difference between 
the nominal and modified cross sections. In addition to the above 
(strongly correlated) 
point-to-point systematics, a systematic of 8.8\% common to all points 
due to uncertainty in the extraction of the integrated luminosity is
quoted, but not 
included in the height of the systematic uncertainty boxes. 

The theoretical cross section was corrected for underlying event and 
hadronization (UEH) effects. The dominant contribution from 
the UEH to the di-jet mass is from the individual jet masses \cite{Proceedings:2016tff}, which are 
typically treated as massless in NLO calculations. 
The UEH correction was estimated from simulation by 
taking the ratio of the particle-level over parton-level  di-jet yields. The 
ratio ranges from 1.44 at low mass to 1.22 at high mass and is used as a 
multiplicative correction to the 
NLO predictions. 

The systematic uncertainty 
on both the UEH correction (double-hatched red band) and the 
theoretical cross section itself took into account the uncertainty on 
the PDF set 
used as well as sensitivity to the variation of the factorization and 
renormalization scales, which were altered simultaneously by 
factors of 0.5 and 2.0. The factorization and renormalization scales were 
also varied independently between the limits above, but the resulting 
deviation was always less than the simultaneous case. The systematic 
uncertainty  on the UEH correction ranged between 39\% and 7\% from 
low to high mass, respectively, while the uncertainty on the 
theory was between 19\% and 43\%. The height of the blue hatched band 
represents 
the quadrature sum of the theoretical and UEH systematics. Note that 
neither systematic uncertainty is symmetric about its nominal value.
Systematic uncertainties on the extracted cross section are smaller than 
the theoretical uncertainties for all mass bins, meaning these data have 
the potential to improve our understanding of UEH effects (at low mass) 
and unpolarized PDFs in our kinematic regime. 

Sorting the yields by beam spin state enables
a determination of the longitudinal double-spin asymmetry $A_{LL}$, 
evaluated as
\begin{equation} \label{eq:ALLEXP}
A_{LL} = \frac{\sum \left( P_{Y}P_{B} \right) \left( N^{++} - r N^{+-} \right)}{\sum \left( P_{Y}P_{B} \right)^2 \left( N^{++} + r N^{+-} \right)},
\end{equation}

\noindent where $P_{Y,B}$ are the polarizations of the yellow and blue beams, 
$N^{++}$ and $N^{+-}$ are the di-jet yields from beam bunches with the same 
and opposite helicity configurations, respectively, and $r$ is the relative 
luminosity of these configurations. The sum is over individual runs, 
which ranged from 10 to 60 minutes in length and were short 
compared to changes in beam conditions. The factor $r$ was close to unity 
on average, varying between 0.8 and 1.2.

\begin{figure}[t!]
\centering \includegraphics[keepaspectratio=true,
width=0.99\linewidth]{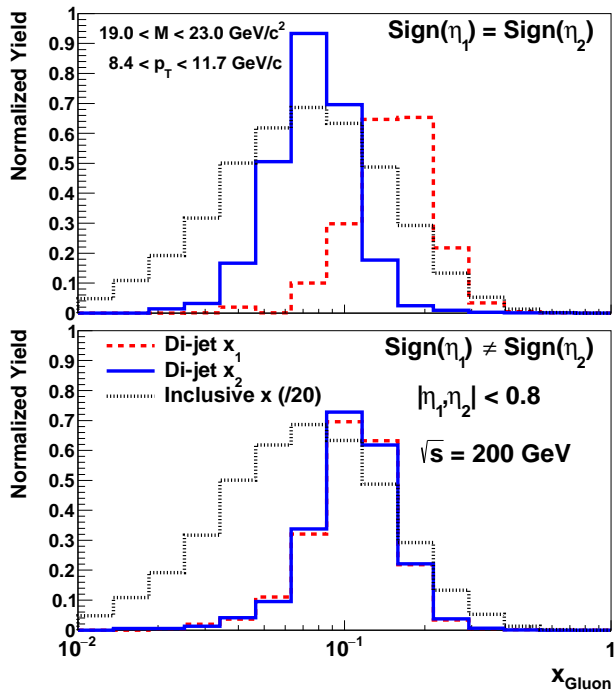}
\caption[Kinematics]{(Color online) Values of gluon $x_{1}$ and $x_{2}$ obtained from the 
PYTHIA detector level simulation for the same-sign (upper) and opposite-sign (lower) di-jet topologies, compared to the gluon $x$ distribution for inclusive jets scaled by an additional factor of 20 in each panel.}
\label{fig:KIN}
\end{figure}

As noted previously, the advantage of a correlation observable
over inclusive measurements lies in the former's superior ability to 
constrain initial state kinematics based on, for example, invariant mass 
and di-jet topological configurations. 
The asymmetry $A_{LL}$ is presented for two distinct 
topologies: `same-sign' in which both jets have either positive or 
negative pseudorapidity, and `opposite-sign' in which one jet has positive 
and the other negative pseudorapidity.
The opposite-sign topology 
selects events arising from relatively symmetric (in $x$) partonic 
collisions, whereas same-sign events select more asymmetric collisions. 
The most asymmetric, high-$p_T$ collisions are preferentially between a 
high momentum (high $x$ and therefore highly polarized) quark and a low 
momentum gluon.  
The control over initial kinematics achievable with di-jets can be 
seen in Fig. \ref{fig:KIN} which presents the partonic momentum fraction 
distributions (weighted by partonic $A_{LL}$) of the gluons as obtained 
from 
PYTHIA for a sample of detector level di-jets with $19.0 < M < 23.0$~GeV/c$^2$, as well as for inclusive jets with $8.4 < p_{T} < 11.7$~GeV/c. 
The increase in $x$ resolution achievable with di-jets compared to 
inclusive jets is evident from the much narrower di-jet $x$ distributions. 
The asymmetric nature of the collisions in the same-sign events 
(upper plot) can be seen in the separation of the high- and low-$x$  
distributions, whereas the opposite-sign events (lower plot) sample an 
intermediate $x$ range. Other di-jet 
mass bin choices sample different gluon $x$ regions.

Values of $A_{LL}$ extracted from the data via Eq.~\ref{eq:ALLEXP} 
represent an admixture
of the asymmetries produced from the three dominant partonic 
scattering sub-processes: $qq$, $qg$, and $gg$. The STAR trigger is more 
efficient for certain sub-processes \cite{Adamczyk:2015}, altering the 
sub-process fractions in the 
data-set and thereby shifting the measured $A_{LL}$. 
Further distortions can arise due to systematic shifts
caused by the finite resolution of the detector 
coupled with a rapidly falling invariant mass distribution. 
Corrections were applied to the raw $A_{LL}$ values to compensate for these 
effects. A trigger and 
reconstruction bias correction was determined by comparing $A_{LL}$ from 
simulation at the detector and parton levels using several polarized PDFs 
which predict asymmetries that `bracket' the measured 
$A_{LL}$ values. Although PYTHIA does not include parton polarization 
effects, asymmetries could be reproduced via a re-weighting scheme 
in which each event was assigned a weight equal to the partonic asymmetry 
as determined by the hard-scattering kinematics and (un)polarized PDF sets.
The trigger and reconstruction bias correction in each mass bin was 
determined by evaluating 
$\Delta A_{LL} \equiv A_{LL}^{detector} - A_{LL}^{parton}$ for each of 
the selected PDFs, then taking the average of the minimum and maximum 
values found. These corrections to $A_{LL}$ varied from 0.0006 at low 
mass to 0.0048 at high mass. Half of the difference between the minimum and 
maximum 
$\Delta A_{LL}$ was taken as a systematic uncertainty on the correction. 

\begin{figure}[t!]
\centering \includegraphics[keepaspectratio=true,
width=0.99\linewidth]{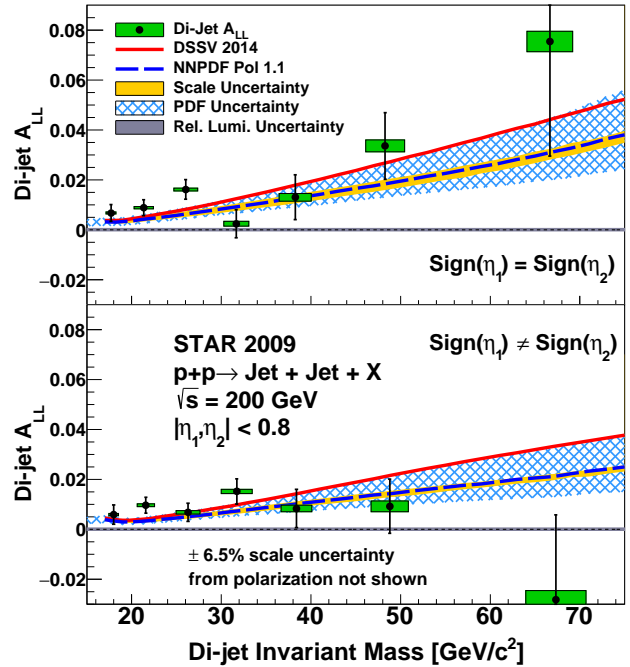}
\caption[Asymmetry]{(Color online) Di-jet $A_{LL}$ vs. parton-level invariant mass for the same-sign (top) and opposite-sign (bottom) topological configurations measured by the STAR experiment. The uncertainty symbols and theoretical curves are explained in the text.}
\label{fig:ALLFIG}
\end{figure}

Figure \ref{fig:ALLFIG} presents the final di-jet $A_{LL}$ measurement 
for the 
same-sign (top) and opposite-sign (bottom) topological configurations as 
a function 
of di-jet invariant mass, which has been corrected back to the parton 
level. 
The correction to parton level is achieved by shifting 
each point by the average difference between the detector 
and parton-level di-jet masses for a given detector-level bin. The 
heights of the uncertainty boxes represent the systematic uncertainty 
on the $A_{LL}$ values due to 
the trigger and reconstruction bias (3--32 $\times 10^{-4}$) and residual 
transverse polarization components in the beams (3--26 $\times 10^{-4}$). 
The 
relative luminosity uncertainty ($5 \times 10^{-4}$) also results in an 
uncertainty in the vertical dimension that is common to all points and is 
represented by the gray band on the horizontal axis.
This uncertainty was evaluated by comparing relative luminosity values 
obtained using the STAR BBCs and ZDCs,  
as well as from 
quantitative inspection of a number of single- and double-spin asymmetries 
expected to yield null results. The widths of the boxes 
represent the 
systematic uncertainty associated with the corrected di-jet mass values and, in 
addition to contributions from the uncertainty on the correction to the parton 
level, include uncertainties on calorimeter tower gains and efficiencies 
as well as TPC momentum resolution and tracking efficiencies. A further 
uncertainty was added in quadrature to account for the difference between 
the PYTHIA parton level and NLO pQCD di-jet cross sections. This PYTHIA 
vs.\ NLO pQCD uncertainty dominates in all but the lowest mass bin, 
rendering the di-jet mass uncertainties highly correlated. The 
$A_{LL}$ values and associated uncertainties can be found in Tab. 
\ref{tab:ALLRESULTSTAB} with 
more detail in the supplemental materials \citep{Supplement}.

\begin{table}[ht!]
\setlength{\tabcolsep}{9.0pt}
\begin{center}
\scalebox{0.9}{
\begin{tabular}{ c  c  c }
\hline \hline
Bin & $\mathrm{Mass} \pm (\mathrm{Sys})$ [GeV/c$^2$] & $A_{LL} \pm (\mathrm{Stat}) \pm (\mathrm{Sys})$ \\ \hline
1 & $17.70 \pm 0.56$ & $0.0067 \pm 0.0034 \pm 0.0004$ \\ 
2 & $21.34 \pm 1.07$ & $0.0088 \pm 0.0032 \pm 0.0005$ \\ 
3 & $26.02 \pm 1.33$ & $0.0162 \pm 0.0039 \pm 0.0006$ \\ 
4 & $31.66 \pm 1.39$ & $0.0024 \pm 0.0056 \pm 0.0010$ \\ 
5 & $38.25 \pm 1.79$ & $0.0130 \pm 0.0089 \pm 0.0015$ \\ 
6 & $48.28 \pm 2.17$ & $0.0336 \pm 0.0133 \pm 0.0024$ \\ 
7 & $66.65 \pm 2.56$ & $0.0755 \pm 0.0460 \pm 0.0041$ \\ 
8 & $17.99 \pm 0.54$ & $0.0059 \pm 0.0039 \pm 0.0005$ \\ 
9 & $21.58 \pm 0.96$ & $0.0096 \pm 0.0032 \pm 0.0006$ \\ 
10 & $26.29 \pm 1.32$ & $0.0068 \pm 0.0037 \pm 0.0007$ \\
11 & $31.72 \pm 1.72$ & $0.0151 \pm 0.0050 \pm 0.0009$ \\ 
12 & $38.38 \pm 1.70$ & $0.0083 \pm 0.0077 \pm 0.0013$ \\ 
13 & $48.79 \pm 2.07$ & $0.0092 \pm 0.0109 \pm 0.0022$ \\ 
14 & $67.32 \pm 3.35$ & $-0.0282 \pm 0.0340 \pm 0.0036$ \\ \hline \hline
\end{tabular}
}
\caption[$A_{LL}$ Values]{Di-jet invariant mass and $A_{LL}$ values with 
associated uncertainties for the same-sign (bins 1-7) and opposite-sign 
(bins 8-14) topologies.}
\label{tab:ALLRESULTSTAB}
\end{center}
\end{table}


Theoretical $A_{LL}$ values were obtained from 
the di-jet production code of de Florian {\it et al.} 
\cite{deFlorian:1998qp} using the DSSV2014 \cite{deFlorian:2014yva} and NNPDFpol1.1 \cite{Nocera:2014gqa} polarized PDF 
sets as input, normalized by the MRST2008 \cite{Martin:2009iq} and NNPDF2.3 \cite{Ball:2012cx} unpolarized sets, respectively. Uncertainty bands representing the sensitivity 
to factorization and renormalization scale (solid) and polarized PDF 
uncertainty (hatched) were generated for the NNPDF result. Overall, the 
data show good agreement with both the DSSV (same-sign $\chi^2/\mathrm{NDF} = 9.9/7$, opposite-sign $\chi^2/\mathrm{NDF} = 9.2/7$) and NNPDF 
(same-sign $\chi^2/\mathrm{NDF} = 12.0/7$, opposite-sign $\chi^2/\mathrm{NDF} = 8.8/7$) predictions. This is 
to be expected as both global analyses incorporated the STAR 2009 
inclusive jet $A_{LL}$ data, of which these results are a subset 
(the correlation matrix between the inclusive and di-jet results 
can be found in the supplemental materials \cite{Supplement}). 
However, for both topological configurations, the measured asymmetries 
tend to lie above the theoretical predictions at low invariant mass. This 
suggests the di-jet data may prefer a somewhat higher gluon polarization 
at low $x$ than the current global analyses. 

The di-jet asymmetry results presented here represent an important 
advance in the experimental investigation of the gluon polarization and 
will be the basis for future high statistics di-jet measurements at STAR. 
Correlation measurements capture a more complete picture of the hard 
scattering 
kinematics and therefore, as shown in Fig. \ref{fig:KIN}, offer better 
determination of the gluon momentum fraction than is possible with 
inclusive 
jet measurements. This improvement in $x$ resolution will allow global 
analyses 
to constrain better the behavior of $\Delta g(x)$ as a function of $x$, 
thus reducing the uncertainty on extrapolations to poorly measured $x$ 
regions and, ultimately, the integrated value of $\Delta g(x)$.

In summary, we report the first di-jet unpolarized cross section and 
longitudinal double-spin asymmetry measurements from STAR in 
polarized $pp$ collisions at $\sqrt{s} = 200$~GeV. The cross section 
result is consistent with NLO pQCD expectations and has the potential to 
constrain unpolarized PDFs. The $A_{LL}$ 
results support the most recent DSSV and NNPDF NLO global 
analyses, which 
included 2009 RHIC data and found the first non-zero $\Delta G$ value for 
$x > 0.05$, and may indicate a slightly higher gluon polarization at 
lower $x$ values.

We are grateful to M. Stratmann and R. Sassot for useful discussions.  
We thank the RHIC Operations Group and RCF at BNL, the NERSC Center 
at LBNL, the KISTI Center in Korea, and the Open Science Grid 
consortium for providing resources and support. This work was
supported in part by the Office of Nuclear Physics within the 
U.S. DOE Office of Science, the U.S. NSF, the Ministry of Education 
and Science of the Russian Federation, NSFC, CAS, MoST and MoE of 
China, the National Research Foundation of Korea, NCKU (Taiwan), 
GA and MSMT of the Czech Republic, FIAS of Germany, DAE, DST, 
and UGC of India, the National Science Centre of Poland, National
Research Foundation, the Ministry of Science, Education and
Sports of the Republic of Croatia, and RosAtom of Russia.



\end{document}

%% file: STAR-Authors-Feb2017.tex
\affiliation{AGH University of Science and Technology, FPACS, Cracow 30-059, Poland}
\affiliation{Argonne National Laboratory, Argonne, Illinois 60439}
\affiliation{Brookhaven National Laboratory, Upton, New York 11973}
\affiliation{University of California, Berkeley, California 94720}
\affiliation{University of California, Davis, California 95616}
\affiliation{University of California, Los Angeles, California 90095}
\affiliation{Central China Normal University, Wuhan, Hubei 430079}
\affiliation{University of Illinois at Chicago, Chicago, Illinois 60607}
\affiliation{Creighton University, Omaha, Nebraska 68178}
\affiliation{Czech Technical University in Prague, FNSPE, Prague, 115 19, Czech Republic}
\affiliation{Nuclear Physics Institute AS CR, 250 68 Prague, Czech Republic}
\affiliation{Frankfurt Institute for Advanced Studies FIAS, Frankfurt 60438, Germany}
\affiliation{Institute of Physics, Bhubaneswar 751005, India}
\affiliation{Indiana University, Bloomington, Indiana 47408}
\affiliation{Alikhanov Institute for Theoretical and Experimental Physics, Moscow 117218, Russia}
\affiliation{University of Jammu, Jammu 180001, India}
\affiliation{Joint Institute for Nuclear Research, Dubna, 141 980, Russia}
\affiliation{Kent State University, Kent, Ohio 44242}
\affiliation{University of Kentucky, Lexington, Kentucky, 40506-0055}
\affiliation{Lamar University, Physics Department, Beaumont, Texas 77710}
\affiliation{Institute of Modern Physics, Chinese Academy of Sciences, Lanzhou, Gansu 730000}
\affiliation{Lawrence Berkeley National Laboratory, Berkeley, California 94720}
\affiliation{Lehigh University, Bethlehem, PA, 18015}
\affiliation{Max-Planck-Institut fur Physik, Munich 80805, Germany}
\affiliation{Michigan State University, East Lansing, Michigan 48824}
\affiliation{National Research Nuclear University MEPhI, Moscow 115409, Russia}
\affiliation{National Institute of Science Education and Research, Bhubaneswar 751005, India}
\affiliation{National Cheng Kung University, Tainan 70101 }
\affiliation{Ohio State University, Columbus, Ohio 43210}
\affiliation{Institute of Nuclear Physics PAN, Cracow 31-342, Poland}
\affiliation{Panjab University, Chandigarh 160014, India}
\affiliation{Pennsylvania State University, University Park, Pennsylvania 16802}
\affiliation{Institute of High Energy Physics, Protvino 142281, Russia}
\affiliation{Purdue University, West Lafayette, Indiana 47907}
\affiliation{Pusan National University, Pusan 46241, Korea}
\affiliation{Rice University, Houston, Texas 77251}
\affiliation{University of Science and Technology of China, Hefei, Anhui 230026}
\affiliation{Shandong University, Jinan, Shandong 250100}
\affiliation{Shanghai Institute of Applied Physics, Chinese Academy of Sciences, Shanghai 201800}
\affiliation{State University Of New York, Stony Brook, NY 11794}
\affiliation{Temple University, Philadelphia, Pennsylvania 19122}
\affiliation{Texas A\&M University, College Station, Texas 77843}
\affiliation{University of Texas, Austin, Texas 78712}
\affiliation{University of Houston, Houston, Texas 77204}
\affiliation{Tsinghua University, Beijing 100084}
\affiliation{University of Tsukuba, Tsukuba, Ibaraki, Japan,}
\affiliation{Southern Connecticut State University, New Haven, CT, 06515}
\affiliation{United States Naval Academy, Annapolis, Maryland, 21402}
\affiliation{Valparaiso University, Valparaiso, Indiana 46383}
\affiliation{Variable Energy Cyclotron Centre, Kolkata 700064, India}
\affiliation{Warsaw University of Technology, Warsaw 00-661, Poland}
\affiliation{Wayne State University, Detroit, Michigan 48201}
\affiliation{World Laboratory for Cosmology and Particle Physics (WLCAPP), Cairo 11571, Egypt}
\affiliation{Yale University, New Haven, Connecticut 06520}

\author{L.~Adamczyk}\affiliation{AGH University of Science and Technology, FPACS, Cracow 30-059, Poland}
\author{J.~K.~Adkins}\affiliation{University of Kentucky, Lexington, Kentucky, 40506-0055}
\author{G.~Agakishiev}\affiliation{Joint Institute for Nuclear Research, Dubna, 141 980, Russia}
\author{M.~M.~Aggarwal}\affiliation{Panjab University, Chandigarh 160014, India}
\author{Z.~Ahammed}\affiliation{Variable Energy Cyclotron Centre, Kolkata 700064, India}
\author{N.~N.~Ajitanand}\affiliation{State University Of New York, Stony Brook, NY 11794}
\author{I.~Alekseev}\affiliation{Alikhanov Institute for Theoretical and Experimental Physics, Moscow 117218, Russia}\affiliation{National Research Nuclear University MEPhI, Moscow 115409, Russia}
\author{D.~M.~Anderson}\affiliation{Texas A\&M University, College Station, Texas 77843}
\author{R.~Aoyama}\affiliation{University of Tsukuba, Tsukuba, Ibaraki, Japan,}
\author{A.~Aparin}\affiliation{Joint Institute for Nuclear Research, Dubna, 141 980, Russia}
% ...

\collaboration{STAR Collaboration}\noaffiliation

%% file: di-jetPaper.bbl
\begin{thebibliography}{50}

\bibitem{Aidala:2012mv} 
  C.~A.~Aidala, S.~D.~Bass, D.~Hasch, and G.~K.~Mallot,
  Rev.\ Mod.\ Phys.\  {\bf 85}, 655 (2013); and references therein.

\bibitem{deFlorian:2008mr} 
  D.~de~Florian, R.~Sassot, M.~Stratmann, and W.~Vogelsang,
  Phys.\ Rev.\ Lett.\  {\bf 101}, 072001 (2008);
  Phys.\ Rev.\ D {\bf 80}, 034030 (2009).

\bibitem{Blumlein:2010rn} 
  J.~Bl{\"u}mlein and H.~B{\"o}ttcher,
  Nucl.\ Phys.\ {\bf B 841}, 205 (2010).

\bibitem{Leader:2010rb} 
  E.~Leader, A.~V.~Sidorov, and D.~B.~Stamenov,
  Phys.\ Rev.\ D {\bf 82}, 114018 (2010).

\bibitem{Ball:2013lla} 
  R.~D.~Ball {\it et al.} [NNPDF Collaboration],
  Nucl.\ Phys.\ {\bf B 874}, 36 (2013).
  
\bibitem{Alekseev:2003sk} 
  I.~Alekseev {\it et al.},
  Nucl.\ Instrum.\ Meth.\ A {\bf 499}, 392 (2003).
  
\bibitem{deFlorian:1998qp}
  D.~de Florian, S.~Frixione, A.~Signer and W.~Vogelsang,
  Nucl.\ Phys.\ B {\bf 539}, 455 (1999)
  
\bibitem{Adolph:2012vj} 
  C.~Adolph {\it et al.} [COMPASS Collaboration],
  Phys.\ Lett.\ B {\bf 718}, 922 (2013)
  
\bibitem{Airapetian:2010ac} 
  A.~Airapetian {\it et al.} [HERMES Collaboration],
  JHEP {\bf 1008}, 130 (2010)
 
\bibitem{Abelev:2006uq} 
  B.~I.~Abelev {\it et al.}  [STAR Collaboration],
  Phys.\ Rev.\ Lett.\  {\bf 97}, 252001 (2006).

\bibitem{Abelev:2007vt} 
  B.~I.~Abelev {\it et al.}  [STAR Collaboration],
  Phys.\ Rev.\ Lett.\  {\bf 100}, 232003 (2008).

\bibitem{Adamczyk:2012qj} 
  L.~Adamczyk {\it et al.}  [STAR Collaboration],
  Phys.\ Rev.\ D {\bf 86}, 032006 (2012).
  
\bibitem{Adamczyk:2015}
  L.~Adamczyk {\it et al.}  [STAR Collaboration],
  Phys.\ Rev.\ Lett.\ {\bf 115}, 092002 (2015).

\bibitem{Adare:2008aa} 
  A.~Adare {\it et al.}  [PHENIX Collaboration],
  Phys.\ Rev.\ Lett.\  {\bf 103}, 012003 (2009).

\bibitem{Adare:2008qb} 
  A.~Adare {\it et al.}  [PHENIX Collaboration],
  Phys.\ Rev.\ D {\bf 79}, 012003 (2009).
  
\bibitem{Adare:2014hsq} 
  A.~Adare {\it et al.} [PHENIX Collaboration],
  Phys.\ Rev.\ D {\bf 90}, 012007 (2014)

\bibitem{deFlorian:2014yva} 
  D.~de Florian, R.~Sassot, M.~Stratmann and W.~Vogelsang,
  Phys.\ Rev.\ Lett.\  {\bf 113}, 012001 (2014)
  
\bibitem{Nocera:2014gqa} 
  E.~R.~Nocera {\it et al.} [NNPDF Collaboration],
  Nucl.\ Phys.\ B {\bf 887}, 276 (2014)
  
\bibitem{Jinnouchi:2004up}
  O.~Jinnouchi {\it et al.},
  arXiv:nucl-ex/0412053.

  
\bibitem{Okada:2005gu} 
  H.~Okada {\it et al.},
  Phys.\ Lett.\ B {\bf 638}, 450 (2006)
  
  \bibitem{CNI:2009}
   B.~Schmidke {\it et al.}, BNL C-A Dept. Rep. C-A/AP/490,
   http://public.bnl.gov/docs/cad/Pages/Home.aspx (2013).

\bibitem{NIM-RHIC}
  K.~H.~Ackermann {\it et al.}  [STAR Collaboration],
  Nucl.\ Instrum.\ Meth.\  A {\bf 499}, 624 (2003), and references therein.

\bibitem{BBC:2005}
 J.~Kiryluk [for the STAR Collaboration],
 arXiv:hep-ex/0501072.

 \bibitem{ANTIKT} 
  M.~Cacciari, G.~P.~Salam, and G.~Soyez, 
  JHEP {\bf 04}, 063 (2008).
 
\bibitem{FASTJET} 
  M.~Cacciari, G.~P.~Salam, and G.~Soyez, 
  Eur.\ Phys.\ J. C {\bf 72}, 1896 (2012).
 
\bibitem{Sjostrand:2006za} 
  T.~Sjostrand, S.~Mrenna, and P.~Z.~Skands,
  JHEP {\bf 05}, 026 (2006).
 
 \bibitem{Perugia_ref}
  P.~Z.~Skands, arXiv:0905.3418.
 
\bibitem{GEANT}
  GEANT 3.21, CERN Program Library.
  
\bibitem{Adye:2011gm} 
  T.~Adye,
  Proceedings of the PHYSTAT 2011 Workshop, CERN, Geneva, Switzerland, January 2011, CERN-2011-006, pp 313-318

\bibitem{Lai:2010vv} 
  H.~L.~Lai, M.~Guzzi, J.~Huston, Z.~Li, P.~M.~Nadolsky, J.~Pumplin and C.-P.~Yuan,
  Phys.\ Rev.\ D {\bf 82}, 074024 (2010)
  
\bibitem{Supplement}
	See Supplemental Material at [URL will be inserted by publisher] for [tables of values and associated systematic uncertainties].

\bibitem{Proceedings:2016tff} 
  H.~Jung, D.~Treleani, M.~Strikman and N.~van Buuren,
  DESY-PROC-2016-01.
  
\bibitem{Martin:2009iq} 
  A.~D.~Martin, W.~J.~Stirling, R.~S.~Thorne and G.~Watt,
  Eur.\ Phys.\ J.\ C {\bf 63}, 189 (2009)
  
\bibitem{Ball:2012cx} 
  R.~D.~Ball {\it et al.} [NNPDF Collaboration],
  Nucl.\ Phys.\ B {\bf 867}, 244 (2013)
  
  
\end{thebibliography}
